\documentclass[iopartl,twocolumn,showpacs,preprintnumbers,amsmath]{revtex4}

\usepackage{graphicx}
\usepackage{dcolumn}
\usepackage{bm}
\usepackage{color}

\begin{document}
\newcommand{\js}[1]
{
\textcolour{red}{{#1}}
}


\title{Particle tracking around surface nanobubbles}
\author{Erik Dietrich$^{1,2}$}
\email{e.dietrich@utwente.nl}
\author{Harold J. W. Zandvliet$^2$}
\email{h.j.w.zandvliet@utwente.nl}
\author{Detlef Lohse$^{1}$}
\email{d.lohse@utwente.nl}
\author{James R. T. Seddon$^1$}
\email{j.r.t.seddon@utwente.nl}
\affiliation{$^1$Physics of Fluids and $^2$Physics of Interfaces and Nanomaterials,  MESA+ Institute for Nanotechnology, University of Twente, P.O. Box 217, 7500 AE Enschede, The Netherlands}

\begin{abstract}
 The exceptionally long lifetime of surface nanobubbles remains one of the biggest questions in the field. One of the proposed mechanisms for the stability is the \emph{dynamic equilibrium} model, which describes a constant flux of gas in and out of the bubble. Here, we describe results from particle tracking experiments to measure this flow. The results are analysed by measuring the Vorono\"i cell size distribution, the diffusion, and speed of the tracer particles. We show that there is no detectable difference in the movement of particles above nanobubble-laden surfaces, as compared to nanobubble-free surfaces.
\end{abstract}
\maketitle

\section{Introduction}

Surface nanobubbles are gaseous domains of nanoscopic size, found on immersed substrates. The first indication for their presence was discovered almost two decades ago \cite{Parker1994}\cite{Miller1999} and the topic has since then grown exponentially \cite{craig2010,seddon2011}. One of the properties of surface nanobubbles which still puzzles the community is their extraordinary long lifetime, causing bubbles to remain stable \cite{Zhang2008}. According to classical expectations, the small radius of curvature (usually microns or less), combined with the high surface tension of the water-gas interface, gives rise to a high pressure inside the bubble, which should quickly drive the gas into solution. Different mechanisms have been considered to explain their stability, including diffusion limitation by surfactants \cite{Ducker2009}, formation of water structures at the interface \cite{Sloan2003}, or even the probability that nanobubbles are totally filled with contaminants of some kind \cite{Evans2004}. 
Over the years, many experiments have been reported indicating that nanobubbles are indeed gas-filled, including different spectrographic methods such as Fourier transform infrared spectroscopy (FTIR) \cite{Miller1999}, and atenuated total reflectance infrared spectroscopy (ATR-IR) \cite{Zhang2007}, \cite{Zhang2008}, and the use of electrolysis to produce surface nanobubbles on HOPG (\cite{Zhang2006}, \cite{Yang2009}, \cite{Hui2009}). 

In 2011, we argued that the peculiar contact angle and limited height of surface nanobubbles would mean that the gas inside must be of Knudsen type \cite{seddon2011c}.
This observation supports the dynamic equilibrium mechanism, proposed three years earlier by Brenner and Lohse \cite{Brenner2008}: The fact that the gas is of Knudsen type means that the motion of the gas molecules is not random, but directed mainly perpendicular to the substrate. Momentum transfer between the gas molecules and water at the bubble interface will result in a recirculation of liquid around the bubble. Gas that dissolves in the liquid would then recirculate around the bubble, allowing re-entry either directly through the three-phase line, or through adsorption at the substrate and surface diffusion.

Our previous paper, where indeed an upward flow is reported, has led to a new point of view for the field. In a recent paper,  in contrast to this idea, Chan and Ohl recently combined optical visualisation of surface nanobubbles with particle tracking \cite{Ohl2012}. Their motion analysis of $200\,nm$ sized tracer particles did not show any recirculation of the liquid and particles near the nanobubbles. However, as the authors explain, their technique is only capable of visualising the largest nanobubbles, with diameters $\geq\,230\,nm$. Taking this into account, one must consider the possibility that the motion of the particles is influenced by a few large nanobubbles, and many nanobubbles with sizes below the optical resolution limit.
In this paper, we present particle tracking measurements to analyse the motion of particles at hydrophobised silicon substrates. We compare results of particle tracking on identical substrates, differing only through the coverage of surface nanobubbles. We measure the Vorono\"i cell characteristics, diffusion coefficients, and particle tracks to address the question: Is there a recirculation near surface nanobubbles?

Similar to Chan and Ohl, we use micro-particle tracking to try to detect recirculation. For this purpose, $1\mu m $ diameter polystyrene particles are used (Fluoro-Max, Thermo scientific, Fremont CA).
These particles are almost neutrally buoyant (density ratio of $\approx\,1.02$) and  small enough to  follow the moving liquid (Stokes number $\approx\,10^{-1}$, assuming a recirculation flow with characteristic size comparable to the bubble radius (microns), and circulation speeds in order of meters per second, as measured earlier \cite{seddon2011c}), yet, they are large enough to be imaged with a microscope, and recorded by a CCD camera. 

\section{Samples and preparation} As substrates, silicon wafers with a native oxide layer (thickness $\approx \,9 nm$) were hydrophobised with PFDTS (per-fluoro-dimethyl-trichloro-siloxane). Hydrophobised silicon is often used in nanobubble research, see for example \cite{Lou2000}, \cite{Tyrrell2001} and \cite{Holmberg2003}. Another substrate commonly used, is Highly Oriented Pyrolytic Graphite (HOPG), \cite{Kameda2008} \cite{Yang2008} but since the silicon produced the best uniform illumination in the microscope, it became our substrate of choice. Nonetheless, a few experiments were conducted using HOPG, which produced results similar to those that will be presented below.

Vapour deposition of the PFDTS monolayer on the silicon wafer was done in an evacuated chamber. The chamber was successively opened to PFDTS and water reservoirs, to introduce the vapours and start the monolayer formation \cite{blabla}.
The advancing and receding contact angles for a water droplet on this substrate were found to be $116$ and $97$ degrees, respectively, as measured by an OCA 15+ apparatus (Dataphysics, Germany).

\begin{figure}
\begin{center}
\includegraphics[angle=0,width=8cm]{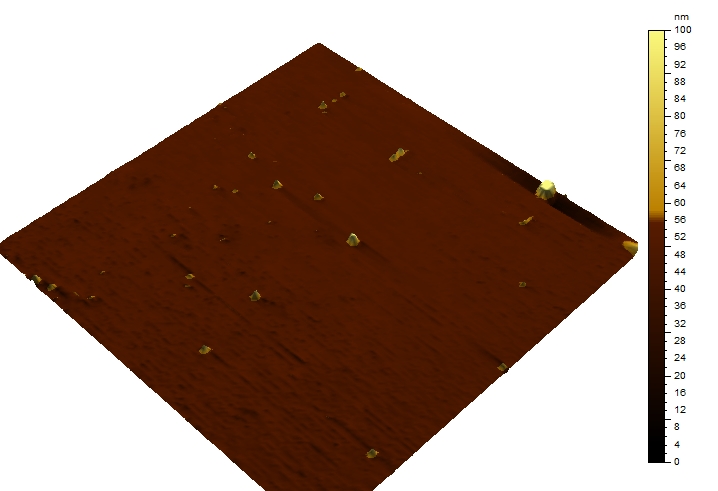}
\end{center}
\caption{\label{fig:AFM} (colour online) Result of an ethanol-water exchange on hydrophobised silicon, as measured by AFM. Field of view is $20\times20\,\mu m^2$ }
\end{figure}

The silicon substrates were cleaned in nitric acid prior to the vapour deposition of the PFDTS. The samples (wafers were cut to small pieces of about $2\,\times\,2\, cm^2$) were ultrasonically cleaned  in isopropyl alcohol for 5 minutes, rinsed thoroughly with ultra pure water, and dried under a stream of nitrogen, prior to each measurement. The same cleaning procedure was used to clean the liquid cell, in which the sample was mounted.
Two types of experiments were performed: \emph{With} bubbles (``gassy") and \emph{without} bubbles (``degassed"). 
In all gassy measurements, bubbles were nucleated using the ethanol-water exchange. The substrate was initially wetted with ethanol (analysis grade, $\geq\, 99.9\%$, purchased from Merck, Germany) whilst fitted in an AFM liquid cell.
Next, during 2 minutes, the $\approx 200\mu L$ of ethanol was gently replaced with $2ml$ of ultra pure water (Millipore Simplicity 185). This method is known to produce nanobubbles \cite{lou2002}, as is illustrated by the AFM image shown in Figure \ref{fig:AFM}. 

\begin{figure}
\begin{center}
\includegraphics[angle=0,width=8cm]{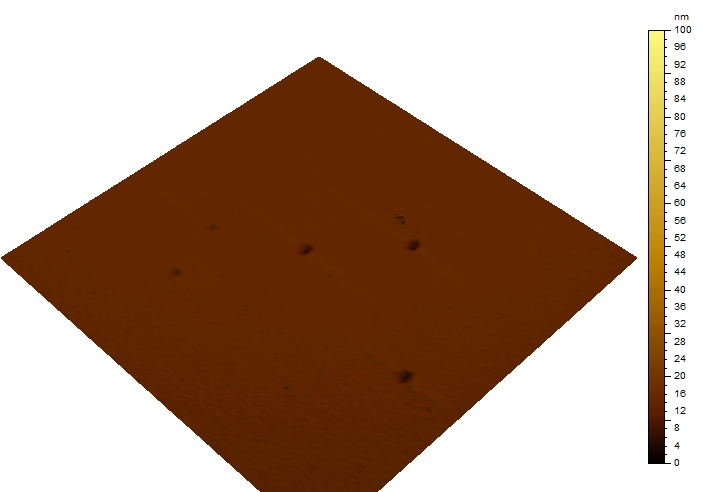}
\end{center}
\caption{\label{fig:AFMdegass} (colour online) Hydrophobised silicon in degassed water, as measured by AFM. Field of view is $15\times15\,\mu m^2$ }
\end{figure}

For all ``degassed" experiments we use degassed water. To this purpose, ultra-pure water was degassed  at a pressure of $1.5\,kPa$ for 1.5 hours, whilst being stirred continuously. The degassed water was then gently poured on the substrate using a syringe. An AFM image of the PFDTS surface under degassed water is shown in Figure \ref{fig:AFMdegass}.

After the preparation of the substrate and liquid, a small volume  ($100\mu L$) of a tracer liquid was added to the liquid cell. The liquid was made by diluting the original particle suspension,containing $1\%$ particles by weight, $1:100$. This diluted suspension was insonicated for $5$ minutes prior to each measurement.

After the particles were added, the sample was placed in the microscope (Olympus, type BX-FM, using a 40x/0.8 water-immersion objective), and the microscope was focussed on the substrate. To do this, a small surface defect was located (usually a small scratch or dent in the silicon toplayer). In this way, only the particles that are in the nearest $\approx 3\mu m$  of the substrate are resolved, resulting in a good surface sensitivity. 
All measurements are conducted in ambient conditions, at room temperature ($21^\circ C\,\pm1\circ$). During the measurement, the sample is shielded by means of a ring, reducing the influence of air convection in the lab.
Once everything was set, the CCD camera (Lumenera LM615, 1.4MPixel monochrome) mounted on top of the microscope recorded the images at a typical rate of 2 frames per second. The combination of the microscope and the camera resulted in a recorded field of view of $155 \times 115\,\mu m^{2}$.
These images were later processed using an in-house developed \textsc{matlab} programme to detect the position of the individual particles.
\newline
\section{Results and Discussion}
We now proceed to describe the results of the various statistical measures that we made to describe the particle positions and motion. 
\newline

\subsection*{Analysis 1: Vorono\"i}

\begin{figure*}
\begin{center}
\includegraphics[trim=0cm 0cm 0cm 0cm,clip,width = 18 cm]{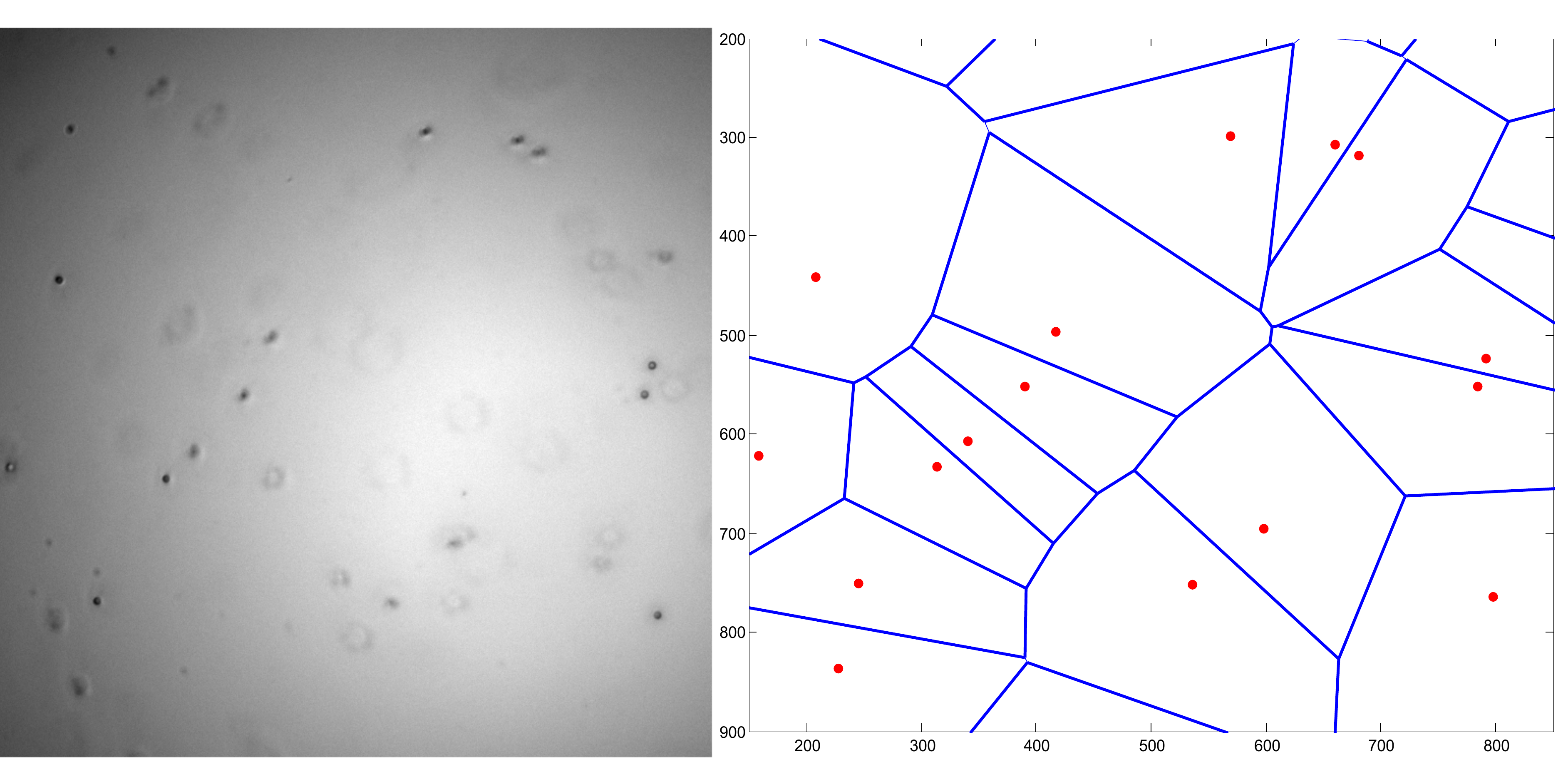}
\end{center}
\caption{\label{fig:Vorodemo_orig} (Left) Outtake of an original frame. The particles are clearly visible as black dots, particles that are out of focus appear as blurred dots, or even as rings and are ignored in the analysis.  (Right) Resulting Vorono\"i cells after analysis of the original image. The centre of masses are shown in red, the blue lines are the cell vertices.}
\end{figure*}

If a strong flow exists around surface nanobubbles, one would expect particles to be ``pushed'' away from the centre, and cluster in-between the bubbles. Clustering and depletion of particles can be characterised by determining the Vorono\"i cell size distribution, as was recently shown by Tagawa and coworkers \cite{Tagawa2012}.
From each recorded frame, the centre of masses of the particles are measured and a Vorono\"i analysis is employed using these points. This routine constructs 2D cells around each centre of mass where each point inside the cell is closest to the corresponding centre. Each vertex of a Vorono\"i cell is constructed by drawing a line perpendicular to, and through the line connecting two neighbouring points. After construction of the Vorono\"i cells, the area of each cell ($A$) is calculated by the programme and stored. We only considered fully-closed cells in our analysis. Some cells, especially near the edges of the frame were ill-defined and not considered in the results. Figure \ref{fig:Vorodemo_orig}  shows a close-up of a single frame, and the resulting Vorono\"i distribution, with the centre of masses in red. 

By calculating the normalised probability density function (PDF) of the Vorono\"i cell sizes, the probability to find cells with a certain size can be compared to a test case, being a reference measurement or theory. We deploy this method by calculating the cell size PDF in the presence of nanobubbles and compare this PDF to the findings in degassed water. As a reference, we compare our measurements to a fitted distribution of a two-dimensional random distribution, as proposed by Ferenc and N\'eda \cite{Ferenc2007}.

When the probability density function of the normalised cell size is calculated, high probability to find small cells is indicative for clustering (small particle-particle spacing) whilst high probabilities for large cells signals depletion.

\begin{figure*}
\begin{center}
\includegraphics[trim=3cm 0cm 0cm 0cm,clip,,width = 20 cm]{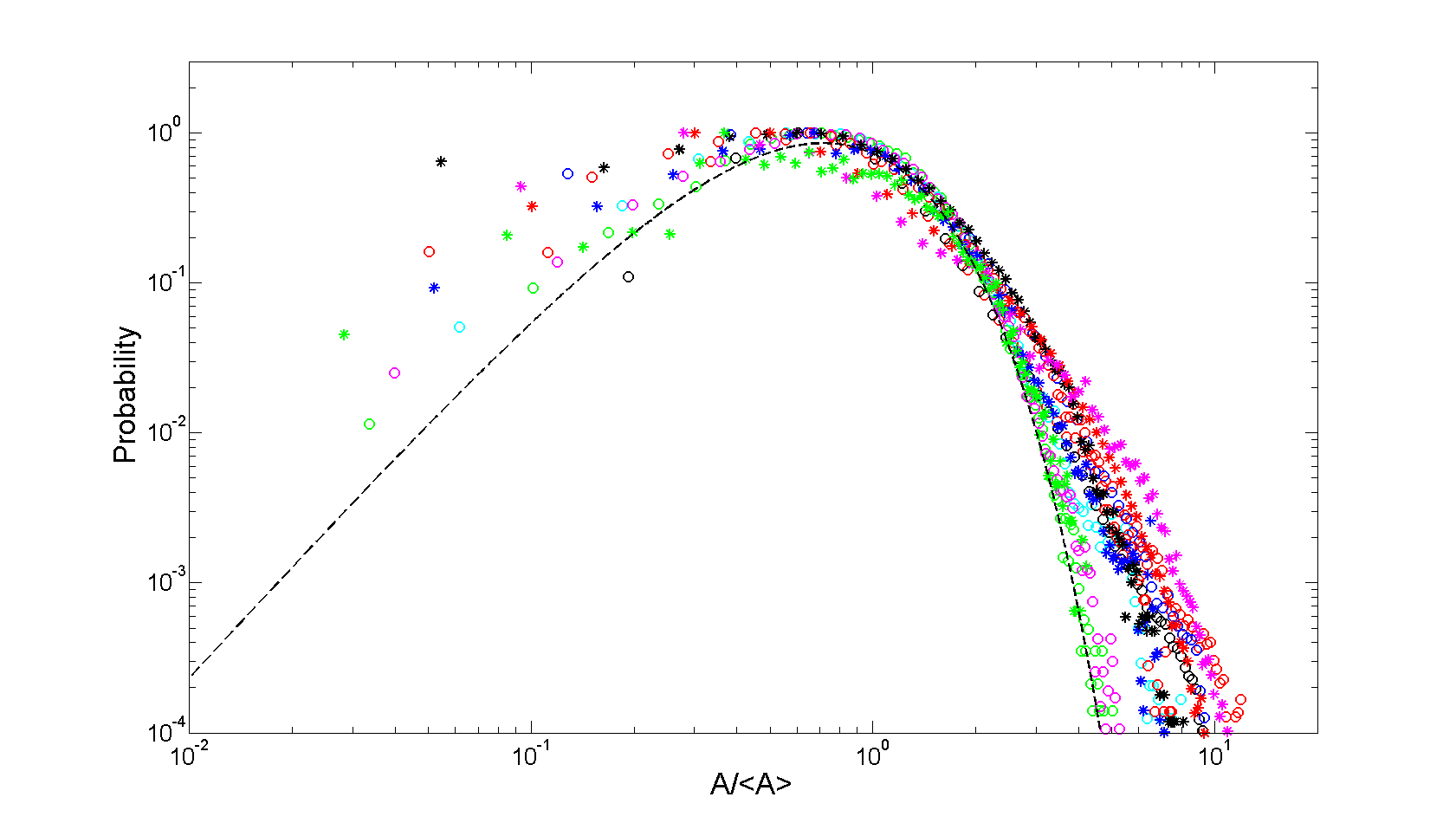}
\end{center}
\caption{\label{fig:voronoi} (colour online) Probability density function of the normalised Vorono\"i cell size. Stars indicate degassed experiments (no bubbles), circles experiments with nanobubbles, the dashed line is the 2D random distribution \cite{Ferenc2007}, colours represent different measurements. When reading the graph, one must realise that the graph is based on a histogram with linear bin sizes, and plotted on a logarithmic scale. }
\end{figure*}

When looking at the right hand side of Figure \ref{fig:voronoi}, the data from the ``degassed'' and the ``bubbly'' experiments coincide nicely, both with respect to each other as well as with the proposed fit. The maxima of all measurements lie within $\pm\,0.3$ of the maximum from the work of Ferenc and Neda \cite{Ferenc2007}. 
On the left hand side of the graph, a deviation from the dashed line is visible for both types of experiments whilst the degassed (starts) and gassy (circles) experiments deviate from one another. Remarkably, the degassed experiments show a higher probability for small $A$, so indicates clustering. It has been shown that dissolved gas influences the dispersion of hydrophobic objects in water \cite{Pashley2005}, it is however beyond the scope of this article to investigate the exact cause of the clustering.
A possible explanation for the difference between the modelled random distribution and experimental data maybe due to the random distribution coming from a purely 2D simulation whilst the experimental images are quasi-2D, accounting for the finite thickness of the slab of liquid imaged by the microscope's small depth of field.

\subsection*{Analysis 2: Diffusion}

If the motion of the particles is purely Brownian, they will display a random walk. If there is a flow near surface nanobubbles, this flow will alter the movement of the particles by either actively pushing them away (above the centre of the bubbles) or trapping them in certain areas. One way to quantify this is by measuring the diffusion of the particles using the mean square displacement. Equation \ref{eq:msd} is the 2D form of the Einstein-Smoluchowski relation, describing the diffusion speed for Brownian motion.

\begin{equation}
<( r{(0)}-r{(t)})^2> =4 Dt
\label{eq:msd}
\end{equation}

\begin{figure}
\begin{center}
\includegraphics[trim=3cm 0cm 3cm 0cm,clip, width= 9cm]{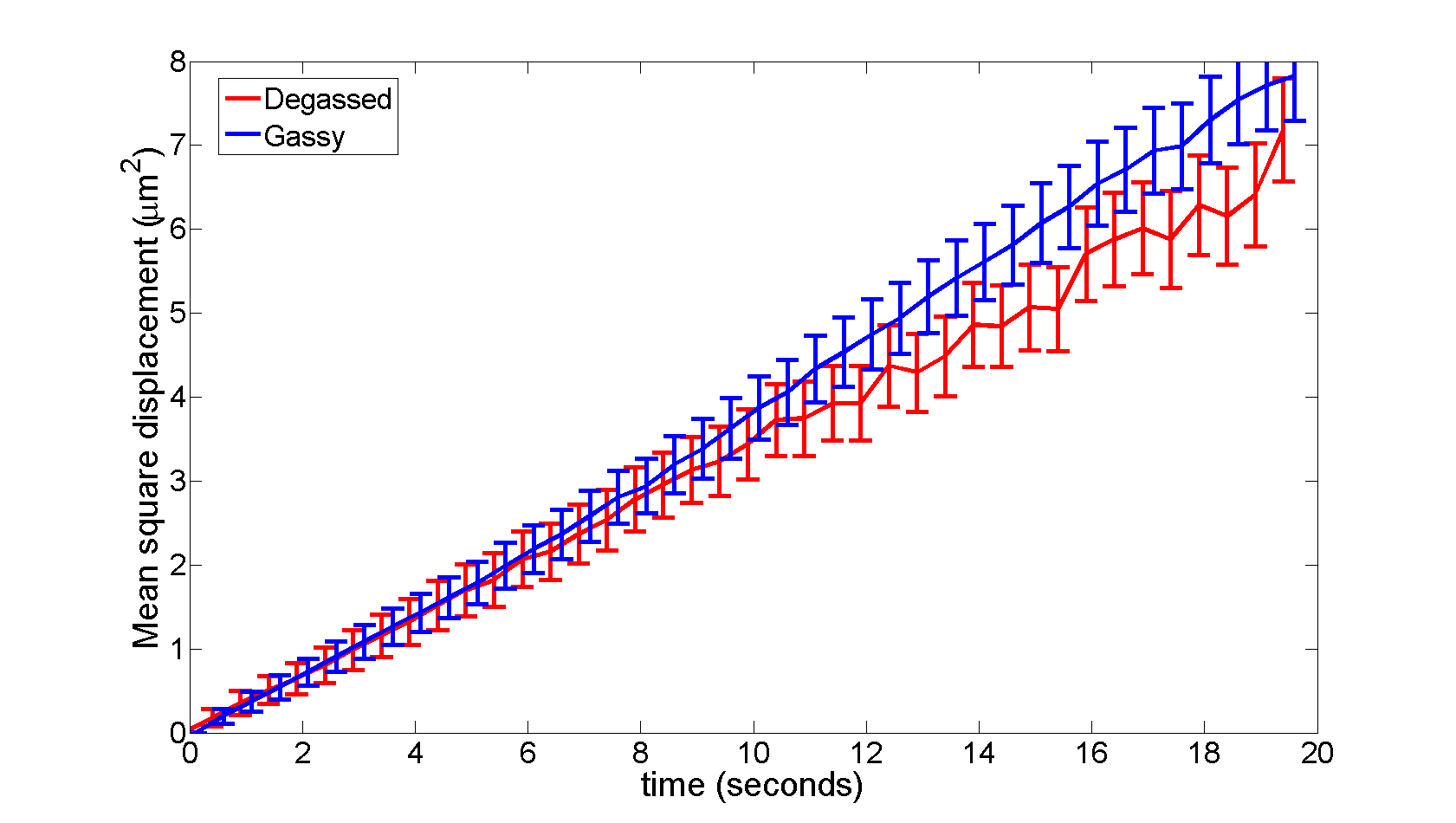}
\end{center}
\caption{(colour online) Mean square displacement as a function of time for all degassed measurements, shown in red, and gassy experiments, shown in blue. }
\label{fig:diffusion} 
\end{figure}

With $r{(0)}$ being the position where the particle is first observed (the starting point), $r{(t)}$ the particle position at time $t$, and $D$ the diffusion coefficient. The pre-factor 4 accounts for the 2D situation, in 3D, it will become 6. If the diffusion is altered by the nanobubbles, we expect to find different values for the diffusion coefficient for the two types of experiments. 
The positions $r{(0)}$ and $r{(t)}$ are calculated using the previously found centre of masses, and a tracking algorithm is used to link the positions in different frames to individual particles. By averaging over all particles, in both gassy and degassed measurements, the mean squared displacement can be plotted as a function of time, as shown in Figure \ref{fig:diffusion}

In Figure \ref{fig:diffusion}, only the first $20$ seconds are shown; the number of particles that stay within the field of view for times larger than this decreases rapidly, resulting in larger errors and deviations. This already becomes apparent in the degassed experiments for $t> 12$ seconds. Again, slight differences appear to emerge between the gassy and degassed experiments, especially at larger time scales. However, the error bar for these times increases steadily, and the discrepancy between the two classes becomes less pronounced.

\subsection*{Analysis 3: Image correlation}

\begin{figure*}
\begin{center}
\includegraphics[trim=3cm 0cm 3cm 1cm,clip, width= 12cm]{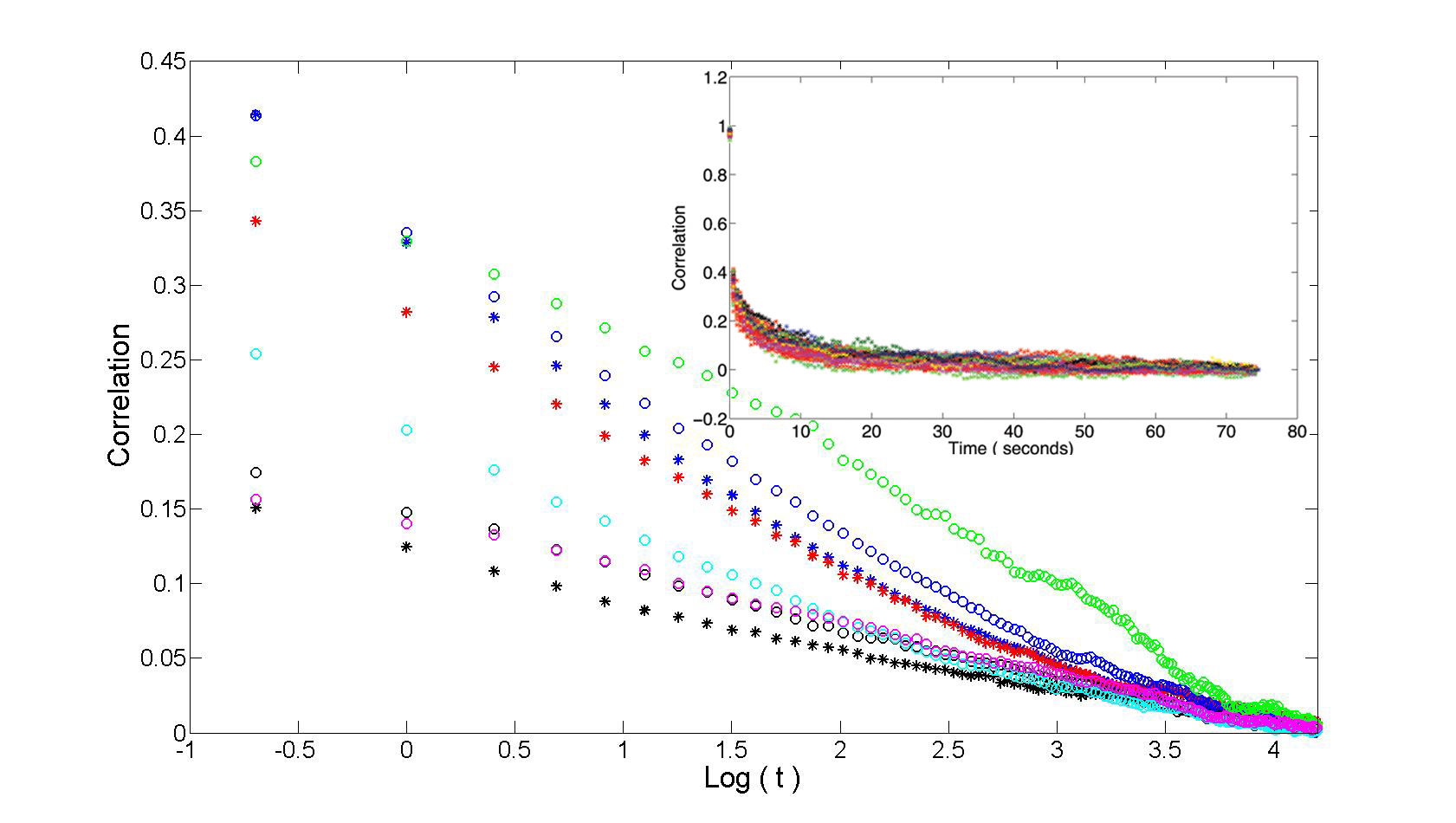}
\end{center}
\caption{(colour online) Correlation coefficient between image taken at $t=0$ and images taken in the subsequent $75$ seconds. Stars indicate degassed experiments, circles gassy experiments. Inset is of one particular measurements, showing different series of correlations in time. }
\label{fig:correlation} 
\end{figure*}

Another common method in microscopy to determine the diffusion speed, is by image correlation: An initial image, taken at $t=0$, is correlated with a series of subsequent images. The correlation coefficient is a measure of how identical the two images are: If imaged particles have moved only slightly, the correlation is large (images are almost identical). The faster the particles move, the quicker the coefficient approaches zero. 
In our experiment the correlation intervals are $75$ seconds. All images in each interval are correlated to the first image of that interval, resulting in an exponential decay of the correlation in time. An example with the results for one measurement is shown in the inset in Figure \ref{fig:correlation}. In Figure \ref{fig:correlation}, the correlation for a series of measurements is plotted versus the logarithm of time, resulting in linear trends. 

The diffusion coefficient is proportional to the slope of this line, so by analysing these slopes, we observe that once again, the spread is considerable. However, there is no clear trend that separates the degassed measurements from the gassy measurements. 

\subsection*{Analysis 4: Local velocities}

\begin{figure*}
\begin{center}
\includegraphics[trim=3cm 0cm 3cm 0cm,clip, width= 18cm]{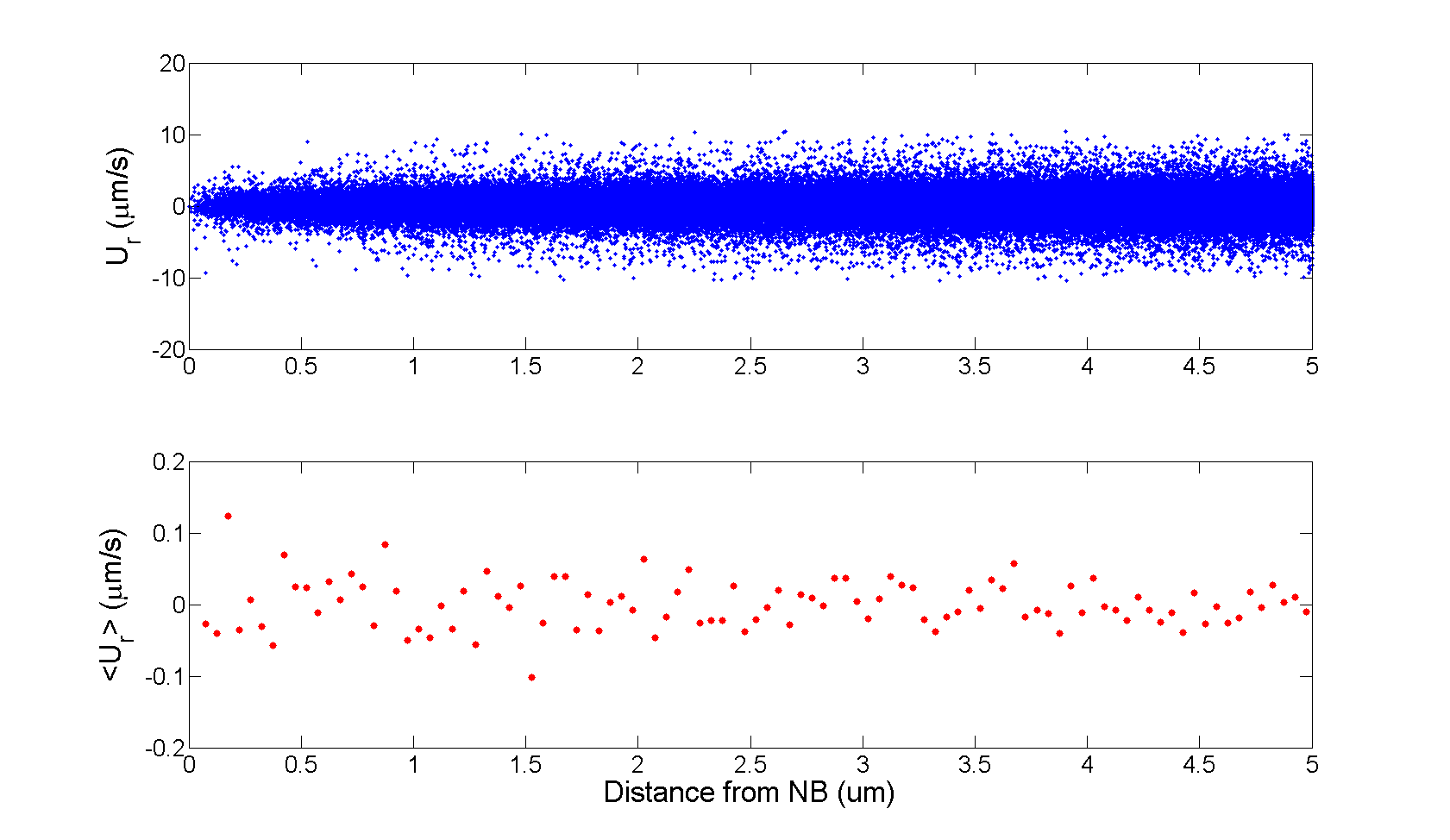}
\end{center}

\caption{(colour online) Radial velocity of particles in the proximity of nanobubbles, as function of their distance from the nanobubbles. Lower graph shows the average value of the radial velocity.}
\label{fig:UofR} 
\end{figure*}

Our final statistical measure is of local particle velocities, following the work of Chan and Ohl \cite{Ohl2012}. We used the information from the particle tracking to construct paths and measure speeds of the particles. The particles did not show any deviation at specific locations, neither was there a noticeable mean velocity. In our efforts described previously \cite{Karpitschka2012}, we employed the interference enhanced reflection microscopy to optically image the nanobubbles. Here as well, we added polystyrene particles and tracked their behaviour. We could not measure any trend in the velocity of particles when they were close to the bubbles. In Figure \ref{fig:UofR} the radial velocity of particles in the proximity of a nanobubble is plotted. It shows that our findings are similar to those reported by Chan and Ohl.

\subsection*{Comment on the validity of the particle tracking technique}

One could question the usability of particle tracking to measure phenomena in the first few microns near a substrate. Electrostatic repulsion between the polystyrene particles and the substrate would result in a minimum spacing between these, below which the repulsion outweighs gravity and van der Waals potential. A quick analysis based on the DLVO theory answers this question\cite{Israelachvili}.
The three potential, in the absence of convection in the liquid, are gravity, van der Waals, and electrostatic potential due to surface charge of the particle and the substrate. 
In this case, gravity is sufficiently small to be ignored, whilst van der Waals potential is calculated with Equation \ref{eq:VdW}:

\begin{equation}
\label{eq:VdW}
F_{VdW}=\frac{-HR}{6D}
\end{equation}

and the electrostatic potential can be calculated using Equation \ref{eq:ED}:

\begin{equation}
\label{eq:ED}
F_{e}=RZe^ { - {\lambda _{D}}^{-1}D }
\end{equation}

With $H$ being the Hamaker constant ($10^{-19}\,J$), $R$ is the radius of the particle ($500\,nm$), $D$ is the distance between the particle and substrate (measured in meters). $\lambda_D$ is the Debye length, in which we assume a dissolved ion concentration of $10^{-7}M$ (pure water), and $Z$ is the reduced surface potential, based on a surface charge of $-80mV$ for the particle and $-56mV$ for the substrate. 

\begin{figure}
\begin{center}
\includegraphics[trim=0cm 0cm 0cm 0cm,clip, width= 8cm]{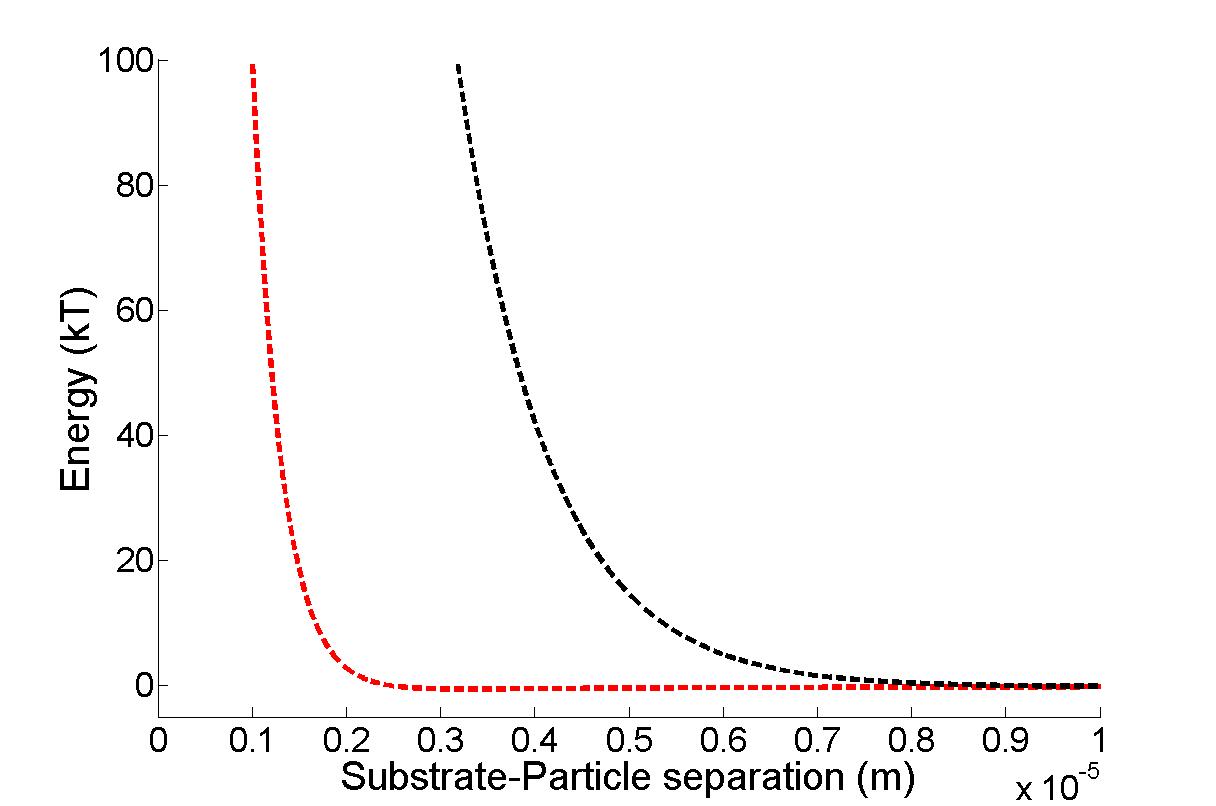}
\end{center}
\caption{Energy as function of substrate-particle separation. Minimum separation will be around $50kT$. Red-dashed line shows calculations for an ion concentration of $10^{-6}M$, black $10^{-7}M$.}
\label{fig:validity} 
\end{figure}

We must consider the depth that particles can penetrate into the repulsive electric field and thus integrate the work needed to bring a particle closer to the wall. In Figure \ref{fig:validity} the combined result of attractive van der Waals and repulsive electrostatics is shown. Assuming that the particle has a kinetic energy around $50kT$, the minimum separation will be around $4\mu m$. As stated above, we assumed an ion concentration of $10^{-7}M$, in practice, small quantities of ions will be dissolved (for example dissolved CO$_2$ gas) which will increase the ionic concentration is increased to $\approx10^{-6}M$. Then the tracer particles-wall minimum separation is reduced to $\approx 1 \mu m$: The tracer particles lie between the $\approx 1.4 \mu m$ minimum separation and the approximately $3$ micron depth of field.

This shows that this method is indeed surface sensitive to within the first few microns near the substrate. Surface sensitivity can be further increased by using particles with a lower surface charge, or by adding salts to the liquid. However, the particle-particle repulsion becomes smaller and clustering of particles might occur.

\section{Conclusion}

In conclusion, our aim was to address the following questions: (i) Is there a statistical variation in flow/diffusion between the nanobubbly and degassed experiments, and (ii) how can we understand the discrepancy between the particle tracking of Chan and Ohl and our recent uplift measurements.
To address the first question: From the Voronoi analysis, see Figure \ref{fig:voronoi}, one could argue that there is a tendency for particles in \textit{degassed} experiments to cluster. This is counter-intuitive, since one would expect that a recirculation flow, if present, would cause clustering of particles in a wide perimeter around the bubble. As becomes apparent from Figure \ref{fig:voronoi}, the spread in the data is considerable, and additional experiments need to be carried out to draw a more solid conclusion.
The same holds for the measurements on the diffusion coefficient. Although there is a spread in the data, as shown in Figure \ref{fig:diffusion}, with a slightly higher diffusion coefficient for the gassy experiments, the results do not indicate a clear difference between gassy and degassed experiments. 

Since we measure (deviations of) the Brownian motion of the tracer particles, our method is highly susceptible to changes in temperature. Also, particles tend to sediment over time and thus increasing the particle density. Thus although the particle density is measured and comparable between measurements, we cannot exclude the possible error due to changing particle numbers.
Our particle velocimetry measurements endorse the conclusions of Chan and Ohl, so based on all results, we too must conclude that, within experimental error, we cannot measure a recirculation flow around surface nanobubbles.

Regarding the second question, our conclusion seems to contradicts findings reported in our earlier work. There, a considerable upflow was measured by moving an AFM tip over a large nanobubble. Here, a $2.7m/s$ upflow was reported, which is in agreement with the proposed Knudsen gas model. One would expect that a flow of this magnitude would be detected easily, and clearly show up in our analysis. The fact that it does not, calls for careful re-analysis of the measurements reported in \cite{seddon2011c}. There is a multitude of explanations for the measured deflection of the AFM tip above a nanobubble, including  changing surface-tip interactions and drift in the vertical stage of the microscope.
However, the Knudsen gas statement still holds, since it only depends on the geometric properties of the nanobubble. As mentioned before, this results in a preferred direction of motion of the gas molecules and the water molecules at the interface. There is however a good possibility that the momentum transfer from the gas molecules to the water molecules is much smaller than previously anticipated. In 2002, de Gennes theoretically described liquid flow over a Knudsen gas film \cite{deGennes2002}, and concluded that slip lengths on the order of microns can be achieved. This would result in liquid velocities at the interface which are much smaller than the predicted values. These small velocities might be undetectable by our method, so a more precise measurement and much larger dataset is needed to test this smaller flow. The presence of a small outflux and diffusion of gas would be in agreement with earlier reports of preferred size distributions for nanobubbles \cite{Borkent2009}\cite{Simonsen2004}, which requires some sort of communication between nanobubbles. 
Recently, alternative mechanisms for nanobubble stability have been proposed by several researchers\cite{Weijs2013,Liu2013,Zhang2013}. These new mechanisms still allow for gas exchange trough the gas-liquid interface, but rely on pinning of the contactline and limitation of the gas outflux to explain the stability.

	Acknowledgements: Initial measurements on the particle motion were conducted as part of the first optical measurements on surface nanobubbles \cite{Karpitschka2012}. The authors are grateful to Stefan Karpitschka and Hans Riegler for providing the opportunity to test the micro particle tracking technique in their lab. 
	D.L., H.Z., J.S. and E.D. acknowledge funding from the Foundation for Fundamental Research on Matter (FOM) and the technology foundation STW, which are sponsored by the Netherlands Organization for Scientific Research (NWO). 

\bibliography{nano_litrev_bib}

\end{document}